\documentclass[aps,prb,twocolumn,groupedaddress,showpacs]{revtex4}%
\usepackage{amsmath}
\usepackage{graphics}
\usepackage{graphicx}
\usepackage{epsfig}
\usepackage{amsmath}
\usepackage{amsfonts}
\usepackage{amssymb}%
\begin{document}
\title{Pomeranchuk effect in unstable $Ytterbium$ systems}
\author{Mucio A. \surname{Continentino}}
\author{Andr\'e S. \surname{Ferreira}}
\affiliation{Instituto de F\'{\i}sica, Universidade Federal Fluminense, \\
Campus da Praia Vermelha, \\
Niter\'oi, RJ, 24.210-340, Brazil.} \email{mucio@if.uff.br}
\date{\today}

\begin{abstract}
$YbInCu_{4}$ and its alloys present discontinuous, first order
iso-structural transitions at  pressure dependent temperatures
$T_V(P)$, where a local moment phase coexist with a renormalized
Fermi liquid phase. We show that along the coexistence line
$T_V(P)$ the entropy of the large volume renormalized Fermi liquid
phase is smaller than that of the higher density, local moment
phase. This implies the existence of a Pomeranchuk effect in these
Kondo lattice materials in analogy with  $^3He$. The theoretical
possibility of using these systems as cooling machines is
discussed.

\end{abstract}
\maketitle

The system $^3He$ presents the unusual feature that along its
melting line, where the solid and liquid phases coexist,  the
entropy of the liquid is smaller than that of the solid
\cite{pomeranchuk,betts}. This occurs for a range of temperature
$T < T_{\times}= 0.32$K, where the melting pressure versus
temperature curve has a minimum, but for $T>T_N$, the temperature
of magnetic ordering of the solid. This feature is the basis of
the cooling technique known as the Pomeranchuk effect
\cite{pomeranchuk} which had an important practical application as
a cooling mechanism to reach very low temperatures in this system.
It is believed to be a unique property of quantum $^3He$ and is a
direct consequence that in its liquid phase this is a renormalized
Fermi liquid, with a linear temperature dependent entropy
$(S_L/Nk_B=3.0T)$ while in the solid phase for the temperature
range above, it may be seen as a collection of weakly interacting
spin-$1/2$ local moments with \cite{pomeranchuk} $S_S/ N k_B
\approx \ln 2$. Also for the low temperatures associated with this
behavior of $^3He$, elastic excitations (the phonons) are quenched
and the relevant degrees of freedom are the magnetic ones due the
spin-$1/2$ nuclei.

\begin{figure}[th]
\begin{center}
\includegraphics[height=5cm]{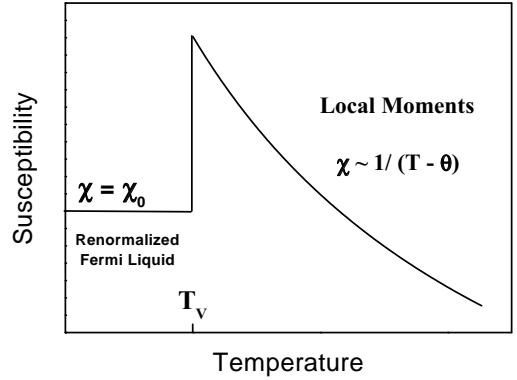}
\end{center}
\caption{Susceptibility curve for $YbInCu_{4}$ (schematic). At
$T_{V} \approx 42$K ($P=0$) there is a discontinuous, first order
iso-structural transition with a volume increase in the low
temperature phase. At $T_{V}$ a local moment phase coexists with a
renormalized Fermi liquid phase \cite{sarrao}. }
\label{fig1}
\end{figure}

The Kondo lattice system $YbInCu_{4}$ presents isostructural
transitions, which for the pure system at ambient pressure ($P
\approx 1$ bar) occurs at a temperature $T_V=42$K \cite{felner,
sarrao}. This is a discontinuous, first order transition which is
accompanied by large changes in transport properties and in
magnetic behavior. In Fig.~\ref{fig1} we show the temperature
dependent magnetic susceptibility of $YbInCu_{4}$ in the range of
the isostructural transition. Above $T_V$, the susceptibility is
of the Curie-Weiss type characteristic of a system of weakly
interacting local moments ($\Theta \approx - 7.2$~K)
\cite{sarrao}. Below the transition, the susceptibility is nearly
temperature independent, i.e., Pauli like as in a Fermi liquid.
The value of $\chi_0$ indicates a renormalized value consistent
with the coefficient of the linear term in the specific heat,
$\gamma =50$ mJ/mol$K^2$ \cite{sarrao}. This leads us to consider
$YbInCu_{4}$ in its low temperature phase as a moderate heavy
fermion system.

This picture of $YbInCu_{4}$ at $T_V$ as a renormalized Fermi liquid phase
coexisting with a local moment phase at a line of first order
transitions $T_V(P)$ brings out a powerful analogy with $^3He$.

Along the melting line $T_{m}(P)$ of $^{3}He$ and the line
$T_{V}(P)$ of the present system, we find coexistence of a
renormalized Fermi liquid phase and a phase of local
moments (in $^{3}He$ the liquid and solid phases, respectively).

The nature of the coexisting phases in $YbInCu_{4}$ leads us to
expect that as in $^3He$, the Fermi liquid phase of this material
has a smaller entropy that of the local moment phase  for some
range of temperatures where these phases coexist.

The Clausius-Clapeyron equation \cite{betts}, as applied to the
$YbInCu_{4}$ system can be written as,
\begin{equation}
\label{eq1}
\left(  \frac{dP}{dT}\right)_{T_V(P)}=\frac{S_{FL}-S_{LM}}{V_{FL}-V_{LM}}%
\end{equation}
where the derivative is obtained at the volume instability curve
$T_V(P)$. The entropy of the local moment phase and of the
renormalized Fermi liquid phase are given by  $S_{LM}$ and
$S_{FL}$, respectively. Notice that, since both phases are truly
solid and the change in volume is small (see Table ~\ref{table}),
the difference $(S_{FL}-S_{LM})$ is mostly due to the magnetic
degrees of freedom. In $^{3}He$  the derivative in Eq.~\ref{eq1}
is calculated at the melting curve where the solid and liquid
coexist. In this system, the melting pressure as a function of
temperature passes through a minimum at $T_{\times}$, where
$S_{solid}=S_{liquid}$ and presents a negative $(dP/dT)$ for
$T<T_{\times}$, consistent with $S_{solid}>S_{liquid}$ in this
range, since the molar volume of the liquid $V_{L}$ is larger than
that of the solid, $V_{S}$ \cite{pomeranchuk,betts}.

\begin{figure}[th]
\begin{center}
\includegraphics[height=6.5cm]{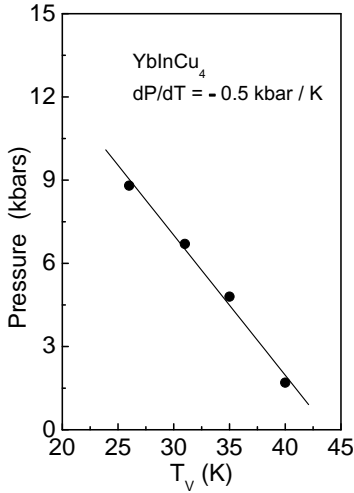}
\end{center}
\caption{ Pressure of the volume instabilities in $YbInCu_{4}$ as
a function of temperature, obtained from resistivity measurements
\cite{sarrao}. }
\label{fig2}
\end{figure}

A plot of the instability temperature for $YbInCu_{4}$ as a
function of pressure $T_V(P)$ (or $P(T_{V})$),  obtained from
resistivity measurements \cite{sarrao} is shown in
Fig.~\ref{fig2}. From the Clausius-Clapeyron equation, the
experimental negative sign of (dP/dT) at $T_V$ and  the fact that
the molar volume in the Fermi liquid phase of $YbInCu_{4}$
($V_{FL}$) is larger than that of the local moment phase $V_{LM}$,
we can conclude that the entropy of the former ($S_{FL}$) at the
coexistence line $P(T_{V})$ is smaller than that of the local
moment phase, for this range of temperatures, just as in $^{3}He$
for $T<T_{\times}$. As pointed out before, the features
$S_{FL}<S_{LM}$, $V_{FL}>V_{{LM}}$  are at the root of the
Pomeranchuk effect in $^{3}He$ which, as we have just shown, also
occurs in the Kondo lattice system $YbInCu_{4}$.
\begin{table}[tbph]
\begin{center}%
\begin{tabular}
[c]{|c|c|c|}\hline & $^{3}He$ & $YbInCu_{4}$\\\hline
Large Volume
Phase (V$_{L})$ & Liquid & Fermi Liquid\\\hline
Small Volume Phase
(V$_{S})$ & Solid & Local Moment\\\hline
$(V_{L}-V_{S})/V_{L}$ &
$0.05$ & $0.005$  \\  \hline $\overline{(dP/dT)}(kbar/K)$ & - $0.016$ & - $0.5$\\\hline
$(W/Q)$ & $>13$ & $0.08$ ($T=40$K)\\\hline
($T_{\times}$, $P_{\times}$) (K, kbar) &
($0.32$, $0.029$) & ($346$, -$151$)\\\hline
\end{tabular}
\end{center}
\caption{Thermodynamic parameters for $^3He$, Ref.~
\protect\cite{betts} and $YbInCu_{4}$, Ref.~
\protect\cite{sarrao}. ($T_{\times}$, $P_{\times}$) for
$YbInCu_{4}$ were obtained extrapolating the Fermi liquid entropy
to reach to non-interacting local moment value.} \label{table}
\end{table}

In Table~\ref{table} we list for comparison some thermodynamic
parameters of $YbInCu_{4}$ and $^{3}He$. Notice that for the
former material, ($T_{\times}$, $P_{\times}$) are obtained from a
crude extrapolation of the Fermi liquid entropy \cite{sarrao}
$S_{FL}/Nk_{B}=6.0 \times 10^{-3}T$  up to that of independent
local moments \cite{gorkov} $S_{LM}/Nk_{B}= \ln (2J+1) =  \ln 8$.

In the case of $^{3}He$ the Pomeranchuk effect is the basis for
the construction of a practical apparatus which has provided the
means for attaining very low temperatures in this system and
eventually discovering its superfluid phases \cite{pomeranchuk}.
The results above show the theoretical possibility of constructing
a Pomeranchuk cooling machine based on the Kondo lattice system
$YbInCu_{4}$. An estimation of the cooling efficiency of such
machine is given by the ratio $(W/Q)$, where
$W=P_V(V_{FL}-V_{LM})$ is the compressional work to squeeze the
Fermi liquid into the local moment phase. The quantity
$Q=T(S_{LM}-S_{FL})$ is the latent heat which represents the
maximum amount of heat which can be removed by converting the
Fermi liquid into the local moment phase. The ratio
$W/Q=-(P_V/T)(dP_V/dT)^{-1}$ attains for $^3He$ its lowest value,
$(W/Q)=13$, at $T=0.14$K \cite{betts}. For $YbInCu_{4}$ this can
be much smaller, for example, $(W/Q)=0.08$ at $T=40$K using the
value of $(dP_V/dT)$ obtained from Fig.~\ref{fig2} (see
Table~\ref{table}).

\begin{figure}[th]
\begin{center}
\includegraphics[height=4cm]{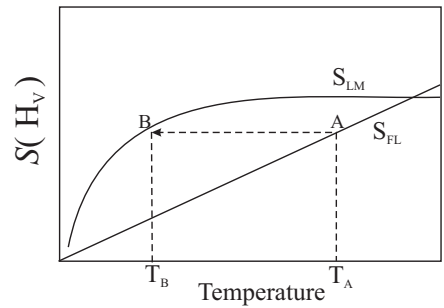}
\end{center}
\caption{ Schematic temperature variation of the entropy of the Fermi liquid and local moment phases at the 
instability field $H_V$. Cooling can be achieved by isentropically increasing the magnetic field to bring the FL
phase into the LM phase as in a process from $A$ to $B$. }
\label{fig3}
\end{figure}

Notice however that instead of adiabatically squeezing
$YbInCu_{4}$ to transform the Fermi liquid into the local moment
phase and reduce the temperature, it is easier to play with the
magnetic field  \cite{sarrao, gorkov} which strongly affects the
volume transition at $T_{V}$. In Fig.~\ref{fig3} we show
schematically the temperature dependence of the entropy $S(H_V,T)$
of the Fermi liquid and local moment phases at the instability
field $H_V$. A similar curve is obtained for $S(P_V,T)$. Cooling
can be achieved by isentropically increasing the magnetic field to
bring the Fermi liquid into the local moment phase, as in the
process $A \rightarrow B$ illustrated in the figure. We point out
that $T_V(H)$ is a universal decreasing function \cite{sarrao} of
$H_V$. Then, the magnetic field can be used as the
\emph{compressing agent} which is simpler than actually squeezing
the Fermi liquid.

In this Letter we have shown the existence of a Pomeranchuk effect
in $YbInCu_{4}$ and its alloys, which opens the theoretical
possibility of using these materials as cooling machines. Our
approach suggests to view the transition at $T_V$ in $YbInCu_{4}$
as the melting of a Kondo lattice.

\begin{acknowledgments}
We would like to thank Conselho Nacional de Desenvolvimento
Cient{\'{\i}}fico e Tecnol\'ogico-CNPq-Brasil
(PRONEX98/MCT-CNPq-0364.00/00), Fundac\~ao de Amparo a Pesquisa do
Estado do Rio de Janeiro-FAPERJ for partial financial support. We
thank J. C. Fernandes and E. Miranda  for useful suggestions and
discussions.
\end{acknowledgments}

\end{document}